\documentclass[twocolumn,showpacs,preprintnumbers,amsmath,amssymb]{revtex4}

\usepackage{graphicx}
\usepackage{dcolumn}
\usepackage{bm}

\begin{document}


\title{Z=50 shell gap near $^{100}$Sn from intermediate-energy Coulomb
excitations in even-mass $^{106 \mbox{--} 112}$Sn isotopes }

\author{C.~Vaman$^1$}
\author{C.~Andreoiu$^2$}
\author{D.~Bazin$^1$}
\author{A.~Becerril$^{1,3}$}
\author{B.A.~Brown$^{1,3}$}
\author{C.M.~Campbell$^{1,3}$}
\author{A.~Chester$^{1,3}$}
\author{J.M.~Cook$^{1,3}$}
\author{D.C.~Dinca$^{1,3}$}
\author{A.~Gade$^{1,3}$}
\author{D.~Galaviz$^1$}
\author{T.~Glasmacher$^{1,3}$}
\author{M.~Hjorth-Jensen$^5$}
\author{M.~Horoi$^6$}
\author{D.~Miller$^{1,3}$}
\author{V.~Moeller$^{1,3}$}
\author{W.F.~Mueller$^{1}$}
\author{A.~Schiller$^{1}$}
\author{K.~Starosta$^{1,3}$}
\author{A.~Stolz$^{1}$}
\author{J.R.~Terry$^{1,3}$}
\author{A.~Volya$^{4}$}
\author{V. Zelevinsky$^{1,3}$}
\author{H.~Zwahlen$^{1,3}$}

\affiliation{ 
$^1$National Superconducting Cyclotron Laboratory, Michigan State
University, East Lansing, Michigan 48824, USA\\
$^2$Department of Physics, University of Guelph, Guelph, Ontario,
Canada N1G 2W1\\
$^3$Department of Physics and Astronomy, Michigan State University,
East Lansing, Michigan 48824, USA\\ 
$^4$Physics Department, Florida State University, Tallahassee,
Florida 32306, USA\\
$^5$ Department of Physics and Center of Mathematics for Applications, 
University of Oslo, N-0316, Oslo, Norway\\
$^6$ Department of Physics, Central Michigan University, Mount Pleasant, Michigan 48859, USA
}

\date{\today}

\begin{abstract}
Rare isotope beams of neutron-deficient $^{106,108,110}$Sn
nuclei from the fragmentation of $^{124}$Xe were employed in an
intermediate-energy Coulomb excitation experiment yielding $B(E2,
0^+_1 \rightarrow 2^+_1)$ transition strengths. The results indicate 
that these $B(E2,0^+_1 \rightarrow 2^+_1)$
values are much larger than predicted by current
state-of-the-art shell model calculations. This discrepancy can be
explained if protons from within the Z = 50 shell are contributing to
the structure of low-energy excited states in this region.
Such contributions imply a breaking of the doubly-magic
$^{100}$Sn core in the light Sn isotopes.
\end{abstract}

\maketitle

Numerous experimental and theoretical studies are currently focused on
nuclear structure evolution far from the line of stability. In
particular, the structure of neutron-deficient nuclei near the N=Z
line is impacted by protons and neutrons occupying the same shell
model orbitals. This letter reports observations which indicate that
large spatial overlaps of valence orbitals in neutron-deficient,
even-mass, tin isotopes, break the stability of the Z=50 shell gap
near doubly-magic $^{100}$Sn.

$^{100}$Sn is the heaviest, doubly-magic, N=Z, particle-bound nucleus
and therefore is of great interest for shell theory of heavy
nuclei. However, it is very difficult to produce and experimentally
study this nucleus.  One way to approach $^{100}$Sn is to examine the
evolution of nuclear properties along the Z = 50 chain of tin
isotopes, which is the longest shell-to-shell chain of semi-magic
nuclei investigated in nuclear structure to date.  The nearly constant
energy of the first excited 2$^+_1$ state between N=52 and N=80\
\cite{ram01}, is one of the well known features of Sn isotopes, and it
seems to indicate that effective nuclear interactions between nucleons
of the same flavor outside a doubly-magic core do not affect the
near-spherical nuclear shape\ \cite{cas01}.  A probe of the stability
of the Z=50 shell gap is provided by the electromagnetic transition
rates between the $0^+_1$ ground and the first excited $2^+_1$ state,
in even mass Sn isotopes. Even small admixtures of proton excitations
across the Z=50 shell gap enhance significantly the electric
quadrupole transition probability between the ground and the first
excited states in contrast to the configurations with the closed Z=50
core and only neutrons in the valence space.

While experimental 2$^+_1$ state energies are well established, the
reduced probability for the electric quadrupole transition from the
ground state to the first excited state, $B(E2, 0^+_1 \rightarrow
2^+_1)$, has been sparsely known except for stable Sn isotopes. For
neutron-rich tin nuclei, the measurements of these $B(E2)$ values have
only recently been achieved due to progress in radioactive beam
techniques\ \cite{rad05}. On the neutron-deficient side, the
corresponding numbers are still unknown except for $^{108}$Sn measured
recently in an intermediate-energy Coulomb excitation at GSI\
\cite{ban05}. The measurements on the neutron-deficient side of the
Z=50 chain are hindered by the 6$^+_1$ isomeric state with a lifetime
in the nanosecond range, while the expected lifetime for the 2$^+_1$
state is at least two orders of magnitude shorter. Therefore, for a
measurement, the $2^+_1$ state must be populated from the ground
state. Consequently, Coulomb excitation is the method of choice if
beams of unstable nuclei are available, while other reactions, in
particular fusion-evaporation, cannot be applied. This letter reports
on the results of an intermediate energy Coulomb excitation experiment
and the measurements of the corresponding $B(E2, 0^+_1\rightarrow
2^+_1 )$ strength of neutron-deficient $^{106-110}$Sn isotopes from
the fragmentation of $^{124}$Xe. In addition, a measurement for
$^{112}$Sn is reported as a check of consistency with existing
experimental data.

Beams of rare isotopes are produced via projectile fragmentation at
the National Superconducting Cyclotron Laboratory (NSCL) as documented
in\ \cite{sto05}. In the current experiment a stable beam of
$^{124}$Xe was accelerated by the Coupled Cyclotron Facility to 140
MeV/nucleon and fragmented on a ~300 mg/cm$^2$ thick Be foil at the
target position of the A1900 fragment separator\ \cite{mor03}. A
combination of slits and a 165~mg/cm$^2$ Al wedge degrader were used
at the A1900 to enhance the purity of the fragment of interest in the
resulting cocktail beam. The properties of the Sn beams in this
experiment are listed in Table\ \ref{A1900beams}.
\begin{table}[ht]
\caption{Properties of the rare isotope Sn beams used in 
the current experiment. }
\begin{ruledtabular}
\begin{tabular}{|c|c|c|c|c|}
Isotope    & Energy  & Purity    & $\Delta$p/p & Rate \\ 
           & [MeV/u] & [\%]      & [\%]        & [10$^3$ pps/pnA]\\
\hline\hline
$^{112}$Sn &   80    &  50       &   0.10      &  19  \\
$^{110}$Sn &   79    &  50       &   0.10      &  21 \\
$^{108}$Sn &   78    &  17       &   0.34      &  17  \\
$^{106}$Sn &   81    &   2       &   0.34      &  0.7 \\
\end{tabular}
\end{ruledtabular}
\label{A1900beams}
\end{table}

Coulomb excitation of the above cocktail beams on a 212~mg/cm$^2$
thick $^{197}$Au target were studied using a combination of the
Segmented Germanium Array (SeGA)\ \cite{mue01} for gamma-ray detection
and the high resolution S800 spectrograph for particle identification
and reconstruction of the reaction kinematics\ \cite{baz03}.  For all
four tin isotopes studied, a lithium-like and a beryllium-like charge
state were delivered to the S800 focal plane and identified by their
position on the Cathode Readout Drift Chamber (CRDC) detectors\
\cite{you99}. The mass and charge of the nuclei were extracted on an
event-by-event basis from the time of flight and energy loss
information.

The S800 CRDC detectors measure position and angle in dispersive and
non-dispersive directions at the focal plane. This information can be
used to reconstruct the trajectories of identified particles to the
target position based on the knowledge of the magnetic field in the
S800\ \cite{baz03}. The proper trajectory reconstruction provides
information on the scattering angle at the target, and, therefore, on
the impact parameter in the Coulomb excitation process\
\cite{win79}. This information is crucial to relate the Coulomb
excitation cross section at the intermediate energies to the reduced
E2 transition probability.  For the projectile excitation this
relation is given by\ \cite{gla99}:
\begin{equation}
\label{sigma}
\sigma_{proj}(E2,I_i\rightarrow I_f)\propto B(E2,I_i \rightarrow I_f)
Z^2_{tar}/b^2_{min},
\end{equation}
where $b_{min}$ is the minimum impact parameter considered for the
cross section measurement. The minimum impact parameter is chosen to
be large enough to minimize the impact of nuclear force interference.

The procedure outlined above for a $B(E2, 0^+_1 \rightarrow 2^+_1)$
measurement from an angle-integrated Coulomb cross section has been
applied in a number of successful experiments at the NSCL\
\cite{gla99, cook01, agade}. In the current study, however, the
absolute Coulomb excitation cross section measurement was hindered by
angular acceptance effects related to properties of the heavy mass and
large charge beams.  Thus, below, the experimental information on the
transition rates was extracted from a relative measurement to
excitations of the $^{197}$Au target.

Following Eq.\ \ref{sigma} the ratio of the cross sections for the Sn
projectile and Au target excitations in the current experiment is
given by:
\begin{eqnarray}
\label{sigmarat}
&&\frac{\sigma_{Sn}(E2,0^+_1\rightarrow 2^+_1)}
     {\sigma_{Au}(E2,3/2^+_1\rightarrow 7/2^+_1)}=\;\;\;\;\;\;\;\;\;\;\;\nonumber\\&&
 \frac{B_{Sn}(E2,0^+_1 \rightarrow 2^+_1)}{B_{Au}
(E2,3/2^+_1 \rightarrow 7/2^+_1)}\left( \frac{79}{50}\right)^2
\end{eqnarray}
The dependence on $b_{min}$ and the reaction kinematics in this ratio
is removed as long as safe Coulomb conditions are met. The ratio of
the cross sections is measured from the ratio of gamma-ray intensities
depopulating the $2^+_1$ state in Sn and the 7/2$^+_1$ state in the Au
nuclei. Knowing the target $B(E2\uparrow)$\ \cite{zho95} the
corresponding transition rate for the projectile is extracted.

\begin{figure}
\includegraphics[angle=0,width=7cm]{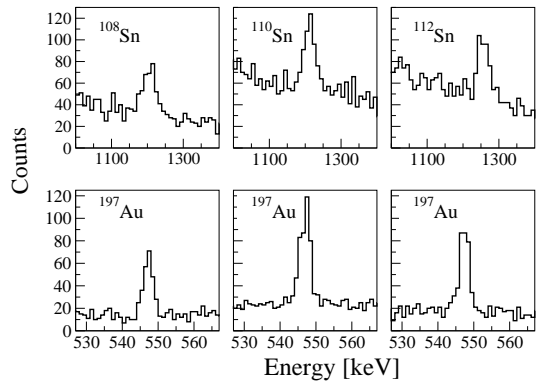}
\caption{\label{spec1} Gamma-ray spectra measured by the 90$^\circ$
ring of the SeGA for the $^{108 \mbox{--}112}$Sn projectiles (top) and
the corresponding Au target (bottom) Coulomb excitations within the 45
mrad scattering angle in the laboratory reference frame.}
\end{figure}

In view of the above, the analysis of the $^{108\mbox{--}112}$Sn data
proceeded in the following way. A subset of particle-identified events
with the impact parameter larger than 19.5 fm was selected; the
corresponding scattering angle in the lab was 45 mrad. Next, the cross
section ratio measurements were performed according to Eq.\
\ref{sigmarat} for the downstream ring at 37$^\circ$ and the upstream
ring at 90$^\circ$ separately, and the $B(E2\uparrow)$'s in Sn nuclei
were extracted from these ratios. Spectra illustrating the quality of
the data for the 90$^\circ$ ring are shown in Fig.\ \ref{spec1}. The
corresponding results are listed in Table \ \ref{res}.  It should be
stressed that the precise value for the impact parameter is not
crucial for this analysis.
\begin{table}[h]
\caption{Reduced E2 transition rates measured for $^{106
\mbox{--}112}$Sn isotopes. The results for $^{108 \mbox{--}112}$Sn
correspond to the lab scattering angles smaller than 45 mrad, for the
$^{106}$Sn the scattering angle limit was set by the S800 spectrograph
acceptance\ \cite{baz03}.
 }
\begin{ruledtabular}
\begin{tabular}{|c|c|c|c|}
Isotope  & $B(E2,0^+_1 \rightarrow 2^+_1)$ [$e^2b^2$]   
& $\Delta_{stat}$ [$e^2b^2$] & $\Delta_{sys}$ [$e^2b^2$]\\
\hline\hline
$^{112}$Sn &   0.240   &  0.020   &   0.025     \\
$^{110}$Sn &   0.240   &  0.020   &   0.025     \\
$^{108}$Sn &   0.230   &  0.030   &   0.025     \\
$^{106}$Sn &   0.240   &  0.050   &   0.030     \\
\end{tabular}
\end{ruledtabular}
\label{res}
\end{table}

\begin{figure}[h]
\includegraphics[angle=0,width=7cm]{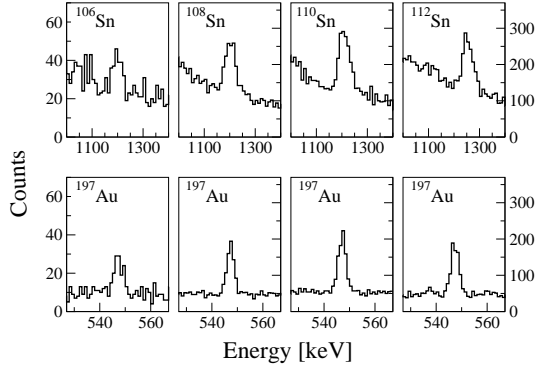}
\caption{\label{spec2} Gamma-ray spectra measured with SeGA for the
$^{106 \mbox{--}112}$Sn projectile (top) and the corresponding target
(bottom) Coulomb excitations within the scattering angle limited by
the S800 spectrograph acceptance.}
\end{figure}

For the $B(E2,0^+_1 \rightarrow 2^+_1)$ measurement in $^{106}$Sn the
off-line analysis requirement set on the impact parameter and the
scattering angle was relaxed; however, the range of the scattering
angles for detected events is still limited to 60 mrad by the angular
acceptance of the S800 spectrograph. For all four isotopes the ratio
of the projectile to the target Coulomb excitations was extracted
using the data shown in Fig.\ \ref{spec2}. A common scaling factor
between these ratios and the measured $B(E2,0^+_1 \rightarrow 2^+_1)$
values was computed for $^{108 \mbox{--}112}$Sn and applied to the
$^{106}$Sn; the resulting $B(E2)$ for $^{106}$Sn is reported in Table\
\ref{res}.
\\

\begin{figure}
\includegraphics[angle=0,width=8cm,height=6cm]{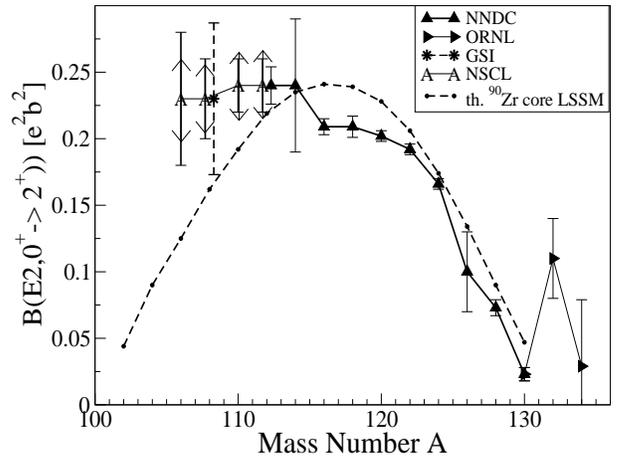}
\caption{\label{resfig} Experimental data on $B(E2,0^+_1 \rightarrow
2^+_1)$ in Sn isotopic chain from the current results for
$^{106\mbox{--}112}$Sn and from Refs.\ \cite{ram01,rad05,ban05}. The
dotted line shows the predictions of the Large Scale Shell Model
calculations of Ref.\ \cite{ban05} performed with $^{90}$Zr core.
For $^{106\mbox{--}112}$Sn the error bars represent statistical errors;
the corresponding systematic errors are marked by arrows. }
\end{figure}

\begin{figure}
\includegraphics[angle=0,width=5cm]{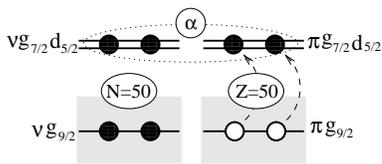}
\caption{\label{smfig} Schematic representation of proton 2p2h
excitations across the Z=50 shell gap in $^{102}$Sn. The occupation of
the same proton and neutron orbitals above the Z=N=50 shell leads to
$\alpha$-like correlations between the valence nucleons. }
\end{figure}

Experimental information on the $B(E2,0^+_1 \rightarrow 2^+_1)$
systematic in Sn isotopes based on the current measurement and Refs.\
\cite{ram01,rad05,ban05} is presented in Fig.\ \ref{resfig}.  The
asymmetric behavior of the $B(E2\uparrow)$ with respect to the N=66
neutron mid-shell at A=116 is striking.  This is in disagreement with
several shell model $B(E2\uparrow)$ predictions including these from
the Large Scale Shell Model calculations of Ref.\ \cite{ban05}
performed with a $^{90}$Zr core, see Fig.\ \ref{smfig} for the
comparison. Shell model calculations consistently predict a
$B(E2\uparrow)$ trend which is nearly parabolic and symmetric with
respect to the midshell\ \cite{ban05,vol05}. It reflects properties of
the even-rank E2 tensor operator in the seniority scheme\
\cite{cas01}.  In regard to other recently proposed theories, the
experimental $B(E2\uparrow)$ strength is underpredicted by the Exact
Pairing model of Ref.\ \cite{vol05}. It should also be pointed out
that while the predictions of Relativistic Quasiparticle Random Phase
Approximation of Ref.\ \cite{ans05} are consistent with the
$B(E2\uparrow)$ values measured here for the most neutron-deficient Sn
isotopes, the overall trend for the Sn isotopic chain in the middle of
the shell is not well reproduced by these calculations.

An effect which can explain large $B(E2\uparrow)$ values in the
neutron-deficient Sn isotopes may arise from correlation energy
associated with nucleons occupying the same orbitals near N=Z line. An
analogy can be drawn between the Sn and Ni isotopic chains.  The
$^{56-78}$Ni isotopes have valence neutron configurations, $
(f_{5/2},p_{3/2},p_{1/2},g_{9/2})^{A-56} $, similar in shell structure
to those of $^{100-132}$Sn, $
(g_{7/2},d_{5/2},d_{3/2},s_{1/2},h_{11/2})^{A-100} $.  Effective
charges take into account coupling between the valence nucleons and
the proton particle-hole excitations of the core not included in the
model space.  The empirical values of $ e_{p}=1.2 $ and $ e_{n}=0.8 $
\cite{ti52} apply to the full $ pf $ shell, and thus take into account
coupling to the 2$\hbar\omega$ giant isoscalar and isovector
quadrupole excitations \cite{brown}.  The $ B(E2\uparrow) $ excitation
strengths obtained in the $ (f_{5/2},p_{3/2},p_{1/2}) $ model with the
GXPF1 interaction \cite{gx1} are 0.0126, 0.0249, 0.0264, 0.0243 and
0.0203 e$^{2}$ b$^{2}$ for $^{58,60,62,64,66}$Ni compared to
experimental values \cite{ram01} of 0.0695(20), 0.0933(15),
0.0890(25), 0.0760(80) and 0.0620(90) e$^{2}$ b$^{2}$, respectively.
The full $ pf $ shell results (including the $ f_{7/2} $ orbit)
obtained with GXPF1 are $ B(E2)=$ 0.065, 0.106, 0.119, 0.082 and 0.047
e$^{2}$b$^{2}$, respectively\cite{ni54}.  The coupling of valence
neutrons to the low-lying $ 1p1h $ proton excitations [$
(f_{5/2},p_{3/2},p_{1/2})(f_{7/2})^{-1} $] of the Ni core could be
taken into account by increasing the neutron effective charge from 0.8
to about 1.1 for all of the Ni isotopes leading to $ B(E2)=$ 0.024,
0.047, 0.050, 0.046 and 0.038 e$^{2}$b$^{2}$.  Thus, in analogy, the
effective charge of $ e_{n}=1 $ used for $^{112-130}$Sn in Ref.\
\cite{ban05} for calculations with the $^{100}$Sn core takes into
account both the low-lying and high-lying (2$\hbar\omega$) quadrupole
vibrations.

But effective charge is not enough to account for the large increase
in the $ B(E2\uparrow) $ value for light Ni isotopes in the full $ pf
$ model space compared to that obtained in the $
f_{5/2},p_{3/2},p_{1/2} $ model space.  To better understand the full
$ pf $ model-space result for $^{58}$Ni we need to consider the type
of two-proton excitations leading to the $4p2h$ configuration shown
schematically in Fig.\ \ref{smfig} for $^{102}$Sn.  The low-lying
spectrum of $^{58}$Ni can be described by mixing of $2p$ and $4p2h$
configurations (relative to $^{56}$Ni) with a collective band
corresponding to the predominantly $4p2h$ state starting at 3.5 MeV
\cite{gx1}. This mixing leads to an enhanced $ B(E2) $ for the ground
state. The excitation energies of the multi-hole states \cite{hor} and
the $B(E2)$ values \cite{now} slowly converge to their full-space
values as a function of the number of nucleons excited from the $
f_{7/2} $ orbital.  The $ 4p2h $ state is low in energy due to the
alpha-correlation energy in the $ 4p $ structure as well as the
pairing energy in the $ 2h $ structure.  The alpha-correlation energy
is particularly large near $ N=Z $ when protons and neutrons are in
the same orbital, and when the valence configuration is ``open" in the
sense that many two-particle couplings are allowed.  As neutrons are
added to $^{56}$Ni, the alpha-correlation energy drops as the
$f_{5/2},p_{3/2},p_{1/2} $ neutron orbitals become filled (and hence
less ``open").  To complete the analogy with Sn, improved results in
comparison to experiment for the middle of the Ni isotopes require the
addition of the $ g_{9/2} $ orbit \cite{lit}.

Thus, by considering these results for the Ni isotopes we can
qualitatively understand (1) the origin of the large neutron effective
charge and (2) a proposed origin for further $ B(E2\uparrow) $
enhancement towards $^{100}$Sn due to $ 2p2h $ proton excitations.  In
analogy to the $ pf $ calculations, we expect the full $ sdg $ model
space results to converge slowly \cite{now} as a function of the
number of nucleons excited out of the $ g_{9/2} $ orbit making the
exact calculation difficult.

The $2p2h$ excitations across the Z=50 shell gap and $\alpha$-like
correlations discussed above also influence observables other than
$B(E2\uparrow)$'s. The correlations are likely to impact the
$\alpha$-decay rates for nuclei above $^{100}$Sn.  Next, low-lying
0$^+$ states in the light Sn isotopes built predominantly on $2p2h$
proton excitations are expected to exist close to the ground state
with collective bands built on top of them.  Last, a smooth band
termination\ \cite{afa99} is expected for these bands due to the
limited valence space. All these can be addressed experimentally.

In summary, the measured nearly constant $B(E2, 0^+_1 \rightarrow
2^+_1)$ strength of $\sim$0.24 $e^2b^2$ in $^{106\mbox{--} 110}$Sn
isotopes is in disagreement with the current state-of-the-art shell
model predictions.  This discrepancy could be explained if protons
from within the Z = 50 shell contribute to the structure of low-energy
excited states in this region. Such contributions are favored and
stabilized by the $\alpha$-like correlations for protons and neutrons
occupying the same shell model orbitals.  This result indicates
breaking of the Z = 50 and N=50 gaps near the doubly-magic $^{100}$Sn.

\begin{acknowledgments}

This work is supported by the US National Science Foundation Grants
No. PHY01-10253, PHY-0555366 and NSF-MRI PHY-0619407.  One author
(C.A.) would like to acknowledge the support received from the
National Science and Engineering Research Council of Canada, the
Swedish Foundation for Higher Education and Research and the Swedish
Research Council.  The authors would like to acknowledge computational
resources provided by the MSU High Performance Computing Center and
Center of High Performance Scientific Computing, at Central Michigan
University.
\end{acknowledgments}


\begin{thebibliography}{99}
\bibitem{ni58}The three $  f_{5/2},p_{3/2},p_{1/2}  $ single-particle energies are readjusted to reproduce the spectrum of $^{57}$Ni. For the $B(E2)$ We use harmonic-oscillator radial wavefunctions with $b=1.01 A^{1/6}$).
\end{thebibliography}
\end{document}